\documentstyle[aps,epsf,preprint]{revtex}

\begin{document}
\draft


\title{Non-Fermi Liquid Behavior in a Disordered Kondo Alloy Model.}

\author{D. R.  Grempel$^{1}$ 
\footnote{Presently at Service de Physique de l'Etat Condens\'{e}, 
CEA-Saclay, 91191 Gif-sur-Yvette, France} 
and M. J. Rozenberg$^2$}

\address{$^1$D\'epartement de Recherche Fondamentale sur la Mati\`ere
Condens\'ee, \\SPSMS,  CEA-Grenoble,  
38054 Grenoble Cedex 9, 
France
\\ $^2$ Departamento de F\'{\i}sica, Facultad de Ciencias Exactas y Naturales, \\ Universidad de Buenos Aires, Ciudad Universitaria, (1428) Buenos Aires, Argentina }

\date{February 10, 1999}
\maketitle 
\widetext 
\begin{abstract} 
\noindent 

We study a mean-field model of a Kondo alloy  using  numerical techniques 
and analytic approximations. In this model, 
 randomly distributed magnetic impurities interact with a band of 
conduction electrons and have a residual RKKY coupling of strength 
$J$. This system has a quantum critical point at  
$J=J_{c} \sim T_{K}^0$, the Kondo scale of the problem. The  $T$ 
dependence 
of the spin susceptibility  near the quantum critical point 
is singular with $\chi(0)-\chi(T) \propto T^{\gamma}$ and non-integer 
$\gamma$. At $J_{c}$, $\gamma = 3/4$. For $J\lesssim J_{c}$ there are 
two crossovers with decreasing $T$, first to $\gamma=3/2$ and then to
$\gamma=2$, the Fermi-liquid value. The  
 dissipative part of the 
 time-dependent susceptibility 
$\chi''(\omega)\propto \omega$ as $\omega \to 0$ except at 
 the quantum critical point where we find $\chi''(\omega) \propto 
\sqrt{\omega}$. 
The characteristic spin-fluctuation energy vanishes at the quantum critical point 
with $\omega_{\rm sf} \sim (1-J/J_{c})$ for $J\lesssim J_{c}$, and 
$\omega_{\rm sf} \propto T^{3/2}$ 
at the critical coupling.
\end{abstract}



\section{Introduction} 
\label{intro}

The understanding of metallic states which do not fit within the 
framework of Fermi-liquid theory is one of the important current 
challenges of 
condensed matter physics \cite{conference}. This issue is relevant to 
a  large class of $f$-electron materials which present anomalies 
in their thermodynamic and transport properties at 
low-temperature \cite{experiments}. 
Two important features characterize the physics of these systems. 
One, is the interaction of the conduction electrons with localized magnetic 
moments {\it via} the  
Kondo coupling. The other, is the inevitable presence of disorder 
due to the alloying process. 
Several models in which   
non-Fermi-liquid (NFL) behavior arises as a consequence of the 
interplay 
between disorder and magnetic interactions have been proposed in the 
literature. 
In the Kondo-disorder model 
of Miranda {\it et al.} \cite{miranda}, randomness in the local
hybridization matrix element between localized and itinerant electrons 
is thought to be at the origin of NFL behavior. In this theory,
the disorder generates a 
broad distribution of Kondo temperatures whose tail extends down to 
$T_K=0$. Therefore, a finite fraction of the localized spins
 remain unquenched  at all temperatures and gives rise to 
  singularities in  the 
thermodynamic and transport properties.

In the metallic spin-glass model \cite{sachdev1,sengupta,sachdev3}, the focus 
is on the consequences of randomness in the RKKY intersite couplings. 
This type of disorder is modeled by including a spin-glass-like exchange 
term in the Hamiltonian.  The system has a 
quantum phase transition when the strength of the 
magnetic interaction $J$ becomes comparable to the Kondo 
temperature 
of the underlying Kondo lattice, $T_{K}^{(0)}$. Beyond this point, 
the ground-state 
is no longer a non-magnetic metal but it exhibits
 long-range spin-glass order. NFL 
behavior results from the power-law 
behavior found in the neighborhood of
the quantum critical point, a scenario that is similar to that 
proposed to 
explain NFL behavior in systems close to ferro- or antiferromagnetic 
instabilities \cite{millis}.

In this paper, we study a mean-field Kondo-alloy model recently 
proposed and 
discussed by Sengupta and Georges  \cite{sengupta}. In their 
paper, these authors did not solve the original Hamiltonian but
 a simpler solvable quantum-rotor problem\cite{sachdev2} that was 
assumed to exhibit the same  
 low-frequency behavior. Here, we solve numerically the Kondo-alloy model using 
 classicaland  quantum Monte Carlo techniques. 
We find a quantum phase transition at 
$J=J_{c}\approx 1.15\  T_{K}^{0}$ where the zero-temperature spin-glass 
susceptibility of the system diverges.
At the critical coupling, the $T$-dependence of the uniform  magnetic 
susceptibility is singular with $\chi(0) - \chi(T) \sim  
T^{3/4}$. This anomalous $T$-dependence is also found above the crossover
 line   $T/J_{c}  \sim (1-J/J_{c})^{2/3}$. Below this line, we still 
 find unconventional behavior but the exponent is different, 
 $\chi(0) - \chi(T)  \sim T^{3/2}$. For $J\ne J_{c}$,
  the normal behavior, 
$\delta \chi \sim  T^{2}$,  is recovered at low 
enough temperature. The numerical results 
for the frequency dependence of the susceptibility are very well described 
over a wide range of temperature and frequencies
by a simple approximate expression 
that we derive from the original model in the strong-coupling limit. 
The strong coupling solution reduces to that of the 
simplified model of Sengupta and Georges \cite{sengupta} 
in the $\omega \to 0$ limit. The spin-fluctuation spectrum is 
Fermi-liquid-like for $\omega \to 0$ everywhere except at $J=J_{c}$. 
We find $\chi''(\omega) \propto \omega $ for $\omega \lesssim 
\omega_{\rm sf}$ where the spin-fluctuation frequency  
  $\omega_{\rm sf}\propto (1-J/J_{c})$ for $J\lesssim  J_{c}$
  and $\omega_{\rm sf} \propto 
T^{3/2}$ at the critical coupling. At the quantum critical point, 
$\chi''(\omega) \propto \sqrt{\omega}$. This implies 
 a slow decay of 
 the spin-spin correlation function, $<S_{z}(t) 
 S_{z}(0)> \sim t^{-3/2}$, that anticipates the appearance of 
 long-range order in the system.

The paper is organized as follows: In Section \ref{model} 
we introduce the
model Hamiltonian and use it to derive
an  effective local action for the spin 
degrees of freedom. 
In Section \ref{method}, we discuss two equivalent fromulations of the 
effective model that are well suited for a  
numerical investigation of the problem. 
 These are based on the formal
equivalence between the Kondo-alloy problem and two other models. The 
first one, is 
 a classical one-dimensional Ising chain with short- and 
long-range 
ferromagnetic interactions, and may be solved by classical 
Monte Carlo simulation. The second model system describes
 a single $S=1/2$ 
quantum spin evolving 
 in the presence of two magnetic 
fields, one that 
is time-dependent and random  in the 
longitudinal direction, and another that is static and fixed
 in the  transverse 
direction. This problem may be solved using a quantum Monte Carlo 
algorithm. 
The results of the simulations are presented in Section 
\ref{results} where we also derive a simple analytical 
approximation which allows for a transparent interpretation 
 of the data. This is followed by a comparison of our results to those 
obtained by other authors.

\section{The model} 
\label{model}

In a disordered Kondo alloy, randomly distributed 
 spins interact with a band of conduction 
 electrons through a local Kondo coupling.  There is also a residual 
 RKKY exchange interaction  
between the spins which is random because of the disorder in their 
positions.
 Many of the systems studied 
experimentally exhibit uniaxial anisotropy as a result of 
strong crystal-field and  spin-orbit effects. Therefore, 
to a first approximation, only the coupling between the 
components of the localized spins along the easy-axis 
needs to be considered.  The simplest model  with these 
characteristics is  a Kondo-lattice model with an additional 
Ising-like random exchange term. The Hamiltonian of the 
model is :

\begin{equation} 
H = -\sum_{i,j,\sigma} t_{ij}  c^{+}_{i\sigma} c_{j\sigma} +
J^{\small K}_{||} \sum_{i} S^{z}_i s^{z}_i + \frac{J^{\small 
K}_{\perp}}{2}
\sum_{i} \left(S^{+}_i s^{-}_i+{\rm h.c.}\right)- \frac{1}{2} 
\sum_{i,j} 
J_{ij}S^{z}_iS^{z}_j.
\label{hamil} 
\end{equation} 
Here,  $\vec{S}_i$ is a localized spin operator at the $i$-th site of 
a 
lattice of size $N$. The creation and 
destruction operators for the conduction electrons are 
 $c^{+}_{i\sigma}$ and $c_{i\sigma}$ and $\vec{s}_i=\small{1/2} 
\sum_{\alpha,\beta} c^{+}_{i\alpha}
\vec{\sigma}_{\alpha,\beta} c_{i\beta}$ is the local electronic spin 
density. The nearest-neighbor  electron hopping integral is 
$t_{ij}=t/\sqrt{z}$
 where $z$ is the connectivity of the 
lattice, and $J^{K}_{||}$ and $J^{\small{K}}_{\perp}$
are the longitudinal and transverse Kondo couplings, 
 respectively. The nearest-neighbor couplings between 
the spins, $J_{ij}$, are 
quenched
random variables for which we assume for simplicity  a 
Gaussian probability distribution 
 with zero mean and variance $\left<J^{2}_{ij}\right> = 
J^{2}/z$. The $z$-dependent normalization of the off-diagonal 
couplings $t_{ij}$ and $J_{ij}$ has been chosen such that 
the results in the $z \to \infty$ limit to be considered 
below are finite \cite{metzner}.

The last term 
on the righthand side of  Eq.\ \ref{hamil} is the well-known 
Sherrington 
Kirkpatrick model  \cite{sk} that has a phase transition to a 
 spin glass  state at $T_{g}^{0}\propto J$. 
 The local Kondo coupling favors screening of the  localized spins by the 
conduction 
 electrons  below a characteristic temperature $T_{K}^{0}$. As a  
 consequence of the competition between these two terms,  
 a spin-glass ground-state is only possible for $J\ge J_{c}\sim T_{K}^{0}$. 
 Therefore, $T_{g}\to 0$ as $J\to J_{c}$ from above and 
the system remains paramagnetic down to zero temperature for $J<J_{c}$.
The point $T=0$, $J=J_{c}$ where the nature of the  ground state of the 
system changes defines the quantum critical point \cite{millis,hertz}. 

We investigate the properties of  
model (\ref{hamil}) near the quantum critical point in the framework 
of a dynamical mean-field theory\cite{rmp}.
In this approach,  
exact in the limit of infinite lattice connectivity, the degrees of 
freedom 
 on any particular  lattice site are isolated and treated 
 exactly, while the  rest of the system  is replaced by an effective 
medium 
 to which the  chosen site is coupled. The properties of the 
effective 
 medium are determined self-consistently from the solution of the 
 single-site problem.  In the limit  
 $z\to\infty$,  the 
 configurational average over the random couplings can be performed 
 explicitly \cite{bm} and the intersite terms in Eq.~(\ref{hamil}) 
 can be eliminated. This reduces the problem to a 
 magnetic impurity embedded in an electronic bath and subject  
to a dynamic magnetic self-interaction \cite{rmp,bm}. 
Ignoring for the moment the anisotropy of the Kondo coupling in 
order to simplify the notation, the  effective action of the single 
site 
problem in the paramagnetic phase may be written as \cite{sengupta} :

\begin{eqnarray} 
{\cal S}_{\rm loc}&&= -\int_0^\beta \int_0^\beta d\tau 
d\tau'c^{+}_{\sigma}(\tau){\cal G}_0^{-1}(\tau-\tau') 
c_{\sigma}(\tau') 
+ J_{K} \int_0^\beta
d\tau \vec{S}(\tau) \cdot \vec{s}(\tau)
\nonumber
\\ &&-\frac{J^2}{2} \int_0^\beta  \int_0^\beta d\tau d\tau' S_z(\tau)
\chi(\tau-\tau') S_z(\tau'),
\label{seff} 
\end{eqnarray}

The functions $\chi(\tau)$ and  
${\cal G}_0(\tau)$, {\it a priori} unknown,  are determined by the 
feedback 
effects of the coupling of the impurity site to rest of the system 
through  a set of self-consistency conditions. Their precise form  
depends upon  the shape of the non-interacting electronic density 
of states ${\cal N}(\epsilon)$ of the lattice \cite{rmp}. In the  
case of a 
semicircular 
density of states, the self-consistency equations acquire a 
particularly 
simple form. We have :

\begin{equation} 
\chi(\tau)=\langle{\cal T} \left(S_z(\tau) S_z(0) \right) 
\rangle_{{\cal 
S}_{\rm loc}},
\label{sc2} 
\end{equation} 
for the magnetic degrees of freedom, and 

\begin{equation} 
{\cal 
G}_0^{-1}(\tau-\tau')=\left(-\frac{\partial}{\partial{\tau}}+\mu\right)
\delta(\tau-\tau') -t^{2}{\cal G}(\tau-\tau'). 
\label{g0loc}
\end{equation}
for the electronic degrees of freedom. In the above equations, 
 ${\cal T}$ is the time-ordering operator along the 
imaginary-time 
axis $0\le \tau \le \beta$, $\mu$ is the chemical 
potential and 

\begin{equation} 
{\cal G}(\tau)=-\langle{\cal T}\left( c(\tau) 
c^{+}(0)\right)\rangle_{{\cal S}_{\rm loc}}. 
\label{sc1}
\end{equation} 

It follows that ${\cal G}$ and $\chi$ are, respectively,  
the exact local electronic Green 
function and the imaginary-time dependent spin susceptibility. 
For general ${\cal N}(\epsilon)$, Eq.\ (\ref{g0loc}) is
replaced by a more complicated  implicit condition 
\cite{georges}.

The solution of this set of coupled self-consistent equations is 
still a very difficult task. It may be argued, 
however, that   
 knowledge of the exact bath Green function is not essential for 
the 
understanding of the low-frequency spin dynamics of the model. 
This follows from a perturbative argument \cite{sengupta}
 that establishes 
that 
the long-time behavior of the exact bath Green function is 
qualitatively 
the same as that of a bath of non-interacting electrons, 
{\it i.e.}, ${\cal G}_0(\tau)\sim 1/\tau$. But the form of the
low-energy effective action for the localized spins is determined precisely 
by the 
asymptotic behavior of the electronic Green function. Therefore, if we  
ignore  Eq.\ (\ref{g0loc}) and fix ${\cal G}_0(i\omega_n) = 
\int_{-\infty}^{\infty}
d\epsilon {\cal N}_0(\epsilon)/\left(i\omega_n+\mu-\epsilon\right)$ 
where 
${\cal N}_0(\epsilon)$ is the unrenormalized density of states, 
 we will still get a qualitatively 
correct description of the low-frequency spin dynamics of the model. 

Further progress can be made  by performing
a Hubbard-Stratonovich transformation that decouples the last term in
Eq.~(\ref{seff}) . Introducing  a set of time-dependent random
fields $\eta(\tau)$ that couple to the spin operators \cite{prl1} , the
partition function of the problem may be rewritten as

\begin{equation} 
Z=\int {\cal D}\eta(\tau) 
\exp\left[-\frac{1}{2}\int_0^\beta\int_0^\beta 
d\tau d\tau'
\eta(\tau)\chi^{-1}(\tau-\tau')\eta(\tau')\right] 
Z_K\left[\eta\right], 
\label{z} 
\end{equation} 
where 

\begin{equation}
Z_K\left[\eta\right]=\int {\cal D}c(\tau){\cal D}c^+(\tau)
\mathop{Tr}\limits_{S_z}\left({\cal T}\exp-{\cal S}_{K}\right),
\label{zk}
\end{equation}
and  

\begin{eqnarray} 
{\cal S}_{K}&&= -\int_0^\beta \int_0^\beta d\tau 
d\tau'c^{+}_{\sigma}(\tau){\cal
G}_0^{-1}(\tau-\tau') c_{\sigma}(\tau') +J_{K}\int_0^\beta d\tau 
\vec{S}(\tau)\cdot \vec{s}(\tau)
 \nonumber
\\ &&- {J} \int_0^\beta\ d\tau  S_z(\tau) \eta(\tau). 
\label{seffk} 
\end{eqnarray}

Eq.~(\ref{seffk}) is the action of a single Kondo impurity in a 
 time-dependent magnetic field $J\eta(\tau)$ in the 
$z$-direction. Within dynamic mean-field theory the partition 
function of 
the Kondo alloy is thus given by the  average over all the 
realizations of the 
random field of the partition function of the modified Kondo problem 
of 
Eq.~(\ref{seffk}).

Equations (\ref{z})-(\ref{seffk}) subject to condition (\ref{sc2})
define the mean-field model of the Kondo alloy. In the next Section 
we 
shall show that this model may be cast in two different forms both of 
which are  well suited for setting up schemes  for the numerical 
solution
 of the problem.

\section{Method} 
\label{method}
\subsection{Formulation of the problem}
\label{formulation}

As we are only interested in the spin dynamics of the system, the 
electronic 
degrees of freedom  in Eq.\ (\ref{z})  may be integrated out. 
In the case of the single-impurity Kondo model this leads to the well 
known Coulomb gas 
representation \cite{ay1} of the partition function of the problem. 
The same is true for the generalized problem of Eq.~(\ref{seffk}) as the 
additional random term  commutes with the
longitudinal part of the Kondo coupling. The Anderson-Yuval 
technique\cite{ay1} may therefore be applied to Eq.~(\ref{zk}). 
After averaging  the resulting 
expression  over the distribution of random fields, Eq.~(\ref{z}) may be rewritten as :

\begin{equation} 
\label{zcg} 
Z_{\rm CG}=\sum_{n=0}^{\infty} \int_0^\beta d\tau_1 
\int_0^{\tau_{1}-\tau_{0}} 
d\tau_2\dots
\int_0^{\tau_{2n-1}-\tau_{0}} d\tau_{2n} 
\left(\frac{J^K_{\perp}}{2}\right)^
{2n}
 \exp\left[\sum_{i<j}\  (-1)^{i+j} \ V(\tau_i-\tau_j)\right],
\end{equation} 
where

\begin{equation}
\frac{\partial^2 V(\tau)}{\partial\tau^2}=2\alpha \left(\frac{\pi}
{\beta}\right)^2 \sin^{-2}
\frac{\pi\tau}{\beta}+J^2\chi(\tau), 
\label{v} 
\end{equation}
and $\tau_0$ is a short time cutoff of the order of the inverse 
bandwidth 
of the electron bath. The coupling constant 
$\alpha=(1+\frac{2\delta}{\pi})^2$ 
where $\delta = -\tan^{-1}\left(\pi\tau_0 J^K_{||}/4\right)$ 
is the phase shift 
for scattering of electrons from a local potential $J^K_{||}/4$.

In Eq.~(\ref{zcg}), $\tau_i$, $i=1,\dots,2n$ are the positions on 
the time axis  
of successive
spin-flips generated by the transverse part of the Kondo coupling and
 the 
function $V(\tau)$ represents the interaction between pairs of 
spin-flips. The first term on the righthand side of Eq.\ (\ref{v}) 
 is familiar from the work on the Kondo model \cite{ay1}. It arises from 
the 
singular response of 
the conduction electron bath to a spin flip on the impurity site. The
second term is characteristic of the alloy model and represents the 
reaction of the rest of the spins  to the local perturbation. 
It is interesting to notice that the partition function of a recently 
studied
extended two-band Hubbard model \cite{sikotliar} can be cast in a form 
equivalent to Eq.~(\ref{zcg}).

Equation (\ref{zcg}) is not yet in a form suitable for computation of 
the
magnetic correlation function. We shall next establish a formal
 equivalence between $Z_{\rm CG}$ and the average partition function 
 of a single quantum $S=1/2$ spin in the presence of a random 
Gaussian time-dependent longitudinal field $\xi(\tau)$ and a static 
transverse  
field $\Gamma$, a problem that can be solved numerically  
using 
the quantum Monte Carlo method of references \onlinecite{prl1} and 
\onlinecite{prl2}.
The partition function of the quantum spin problem $Z_{\rm QS}$ is  :

\begin{eqnarray} 
\label{singlespin} 
Z_{\rm QS}= && \int {\cal D}\xi \exp\!
\left[-{1 \over 2}\!\int_0^\beta\!\int_0^\beta\!d\tau 
d\tau'{\xi}(\tau)
Q^{-1}(\tau,\tau'){\xi}(\tau')\right] \nonumber\\ &&\times{\rm 
\mathop{Tr} \limits_{S_z}
{\cal T}}\exp \left[ \int_0^\beta\!d\tau \left[ \xi(\tau){S_z}(\tau) 
+\Gamma {S_x}(\tau) \right] \right], 
\end{eqnarray} 
where $Q(\tau)$ is the correlation function of the random component of the 
magnetic 
field.

To demonstrate the equivalence of (\ref{zcg}) and (\ref{singlespin}), 
we 
first perform a Trotter decomposition of the time-ordered exponential 
in  Eq.~(\ref{singlespin}),

\begin{equation}
\label{trotter}
{\cal T}\exp \left[ \int_0^\beta\!d\tau \vec{h}(\tau) \cdot 
\vec{S}(\tau)
\right]\sim \prod \limits_{k=1}^M \exp\left[\Delta\tau 
\vec{h}(\tau_k) 
\cdot \vec{S}\right],
\end{equation}
where $\Delta\tau=\beta/M$. We next introduce a complete set of 
intermediate 
states $|\sigma_k \rangle \langle\sigma_k|$ at each imaginary 
time-slice 
$\tau_k$. The matrix elements in the Trotter expansion may be 
evaluated 
using the 
expression

\begin{equation} 
\langle \sigma |\exp\left\{{\Delta \tau \left[ \xi(\tau){S_z}(\tau) 
+\Gamma
{S_x}(\tau) \right]}\right\}|\sigma'\rangle \approx e^{\Delta \tau 
\xi(\tau)
\sigma}
\left[\delta_{\sigma \sigma'}+ \delta_{\sigma \bar{\sigma'}} 
\frac{\Gamma 
\Delta
\tau}{2}+{\cal O}(\Delta\tau^2)\right], 
\end{equation} 
valid in the limit $\Delta\tau \to 0$. 
After averaging over the field
$\xi(\tau)$ and taking the limit $M\to\infty$, the partition function 
of the 
model 
may be expressed as a sum of contributions of individual ``spin 
histories'',
 each of 
these being one of the possible sequences of the eigenvalues 
$\sigma(\tau)=\pm 
1/2$ of the intermediate states. We find :

\begin{eqnarray}
Z_{\rm QS} = && \sum_{n=0}^{\infty} \int {\cal D}\sigma^{(n)} 
\exp\left[
\frac{1}{2} \int_0^{\beta}\int_0^\beta d\tau d\tau'
\sigma^{(n)}(\tau) Q(\tau-\tau')\sigma^{(n)}(\tau') + n\ln \Gamma 
\right], 
\label{altyuval}
\end{eqnarray}
where $\sigma^{(n)}(\tau)$ is a spin history with $n$ spin flips in 
the 
interval $0\le\tau\le\beta$ and the integration is over their 
positions. Eq.~(\ref{zcg}) follows from Eq.~(\ref{altyuval}) by 
integrating twice by parts the first 
term 
in the exponential,   
 provided that we choose $\Gamma=J_{\perp}^{K}$ and 
that we identify $Q(\tau)$ with the righthand side of 
Eq.~(\ref{v}). The original problem has thus been reduced to the 
evaluation of 
the partition function of Eq.~(\ref{singlespin}) 
subject to the condition ~(\ref{sc2}). 

An alternative numerical method may be formulated by taking advantage 
of 
the  asymptotic equivalence\cite{ay2,chakra} between the Coulomb gas 
representation (\ref{zcg}) of the Kondo problem and the partition 
function 
of a 
{\it classical} one-dimensional Ising spin chain with 
nearest-neighbor and 
long-range interactions. This problem may be solved numerically using 
standard classical Monte Carlo techniques as has been recently done 
for 
the single-impurity anisotropic Kondo model in reference 
\onlinecite{chakra}.  It may be shown by a straightforward 
generalization of the 
methods of reference\onlinecite{ay2} that the Ising-chain 
model relevant for our problem is

\begin{equation} 
Z_{\rm I}=\sum_{\{S_i\}} \exp\left[ \sum_{i\le L} K_{\rm NN} S_i
S_{i+1} + \sum_{i<j\le L} K_{LR}(i-j) S_i S_{j}\right],
\label{dyson}
\end{equation}
where the ${S_{i}=\pm 1}$ are Ising variables and the
 number of sites in the 
chain is $L=\beta /\tau_0$. The spin-spin interaction 
consists of a short-range part, $K_{NN}
 \approx -1/2\ln \left(J_K \tau_0/2\right)$, and a 
long-range part given by 

\begin{equation} 
K_{LR}(i-j)={1\over 4}
\left[\frac{2 \alpha \left(\pi 
/N\right)^2}{\sin^2\left[\pi(j-1)/N\right]}+
J^2\tau_0^2 \chi(\tau_{0}|i-j|) \right].
\label{jlr}
\end{equation}

It is worth noticing that, while both of these approaches can be 
used to solve the present  
strongly anisotropic Kondo-alloy model, only the first 
one can be generalized 
to the case of a non-Ising spin-spin interaction.

\subsection{Numerical methods}
\label{numerics}

We have simulated the mean-field Kondo-alloy model using the 
two formulations 
described in the previous section as each has its own 
advantages and drawbacks. In particular, while the systematic error 
introduced by the discretization of the imaginary time is larger for
 the quantum simulations,  statistical fluctuations are far more 
important in the classical case. We have empirically
 found that the latter method is 
more 
accurate for the computation of the static susceptibility at low 
 temperatures whereas the former one gives better results for the 
 overall frequency dependence of the spin correlation function.
 
The numerical procedure used to solve the self-consistent 
problem is as follows : i) an approximation to $\chi(\tau)$ is 
used as input in either Eqs.\ (\ref{v}) or (\ref{jlr}).  ii) the 
spin-spin
correlation function is obtained from a Monte Carlo simulation (see 
below).
 iii) a new
$\chi(\tau)$ is computed from  condition
(\ref{sc2}) and used as a new input in step i). This procedure is 
iterated until successive values of the correlation function differ by less 
than a fixed tolerance level (see below).  This 
takes from four to fifteen iterations depending on the 
temperature and the values of the parameters.

The simulations of  the classical problem defined 
by  Eq.~(\ref{dyson}) have been done using a standard Monte 
Carlo heat-bath algorithm for Ising chains of up to 256 sites. 
The quantum problem of Eq.~(\ref{singlespin}) has been simulated as follows.
The imaginary-time 
axis is discretized in slices of width $\Delta\tau=\beta/M$ and the 
time-ordered 
exponential appearing in  Eq.~(\ref{singlespin}) is approximated  by 
a 
Trotter
product of $M$ factors. The statistical weight of a  given
configuration $\{\xi(\tau)\}$ is thus  expressed in terms of the trace of 
a 
product of $2\times 2$ random matrices. The corresponding 
contribution 
to the spin correlation function $\chi(\tau,\tau')$ is computed by 
inserting two additional $\sigma_{z}$ Pauli matrices at the 
appropriate 
places  in the matrix product.
It is important to choose the parameter $M$ appropriately. If $M$ 
is too small,  
the systematic error introduced by the Trotter approximation is large.
 If $M$ is too large, however, the  algorithm is prone to numerical 
instability. 
 We found  that the choice $\beta \lesssim 0.25 M \tau_{0}$ with 
$M\le 128$ 
 is a satisfactory 
compromise. This sets a lower limit to the temperatures that we can 
simulate, $\tau_{0} T_{\rm min}\approx 0.03$.  The simulation 
is most conveniently done in the space of the Masubara-frequency 
components 
of the field,  $\xi(\omega_{n})=\int_{0}^{\beta}d\tau 
\xi(\tau)\exp(-i\omega_{n}\tau)$ \cite{prl1}. 
These are finite in number as a 
consequence of the discretization of time : $\omega_{n}=2\pi n T$ 
with 
$n=0,\dots,M-1$. In an 
elemental Monte Carlo move,  a change of the complex field 
$\xi (\omega_n)$ for a single  frequency is attempted.  A 
full Monte Carlo step is complete when elemental changes have been 
attempted for all the Matsubara frequencies. 

The precision of the numerical calculations  presented below 
is determined by two factors, namely, the statistical error of   
the Monte Carlo calculation and the 
stopping criterion used in the enforcement the self-consistency 
condition. A typical quantum Monte Carlo run consisted of 
$4\times 10^5$  Monte Carlo steps per time slice. This
 corresponds to an absolute error of the order of  
 $2 \times 10^{-3}$ in $\chi(\tau)$. As we mentioned above, 
 the simulations based on Eq.~(\ref{dyson}) are noisier 
than those  based on Eq.~(\ref{singlespin}) which requires an order
 of magnitude more MC steps  to reach the 
same level of accuracy. The stopping criterion for the 
self-consistency loop was that two successive 
values of the static local susceptibility differed by 
less than 0.5{\%}. This 
is about twice the size of the statistical error. 
On the basis of these figures, we  
estimate that our final results for $\chi_{T}$ are 
accurate to within $1{\%}$.

\section{Results}
\label{results}
We have simulated 
the mean-field Kondo alloy model 
for fixed values of the Kondo couplings and several values of $J$ 
for   $T\ge 0.05\ T^0_K$ where $T^0_K$ is the single-site Kondo 
temperature (see below). The first step in the calculation is the choice 
of the 
 parameters $\alpha$, $J^{K}_{\perp}$ and $\tau_0$  that define the 
underlying 
single-impurity Kondo problem (cf. Eq.~(\ref{zcg})). As the 
low-temperature properties of all antiferromagnetic Kondo models are 
described by the same fixed point, we are free to choose these 
parameters using criteria of  numerical convenience. 
For the particular case
 $\alpha=1/2$ and for all values of $J_{\perp}$, 
the Kondo model is equivalent to a simple exactly  
solvable problem, the 
resonant model \cite{toulouse}. We have made this choice as it 
 provides us with means to 
test our numerical methods by comparison of the Monte Carlo 
results means with the analytical solution. We have taken $\tau_0^{-1}$
 as the unit of energy  and we have arbitrarily
  set $J_{\perp}=3/2 \tau_0^{-1}$.

Fig.~\ref{fig1} shows $\chi(\tau)$ 
as a function of the scaled 
variable $\tau/\beta$ for $J=0$ and  several temperatures. 
The correlation function in imaginary time is real and symmetric around 
 $\tau = \beta/2$ as a 
consequence of time-reversal invariance. Its minimum value 
steadily decreases with decreasing 
temperature. The expected behavior of the zero-temperature dynamic 
susceptibility in the long-time limit is 
$\chi(\tau) \propto \tau^{-2}$ \cite{ay1}. At finite 
temperatures
this expression generalizes  to $\chi(\tau) \sim (\pi/\beta)^{2} 
\sin^{-2}(\pi \tau/\beta)$ \cite{ay3}.
We have fitted our data for $\tau \sim \beta/2$ with the expression

\begin{equation}
\label{asym}
\chi(\tau)= \left(\frac{\pi}{\beta}\right)^{2} \frac{A}
{\sin^{2}\frac{\pi \tau}{\beta}+ \sin^{2}\frac{\pi \tilde{\tau}}
{\beta}},
\end{equation}
where $A$ is a $T$-dependent amplitude and the cutoff $\tilde{\tau}$ 
is of the order of the inverse of the Kondo 
temperature to be defined below. 
The fits,  shown by the  solid lines in the 
figure, are in excellent agreement with the numerical data. 

The static spin susceptibility has been   
computed from the Monte Carlo results using 
the expression \cite{bm} $\chi_{T} = \int_0^\beta d\tau \chi(\tau)$. 
The results thus obtained are shown in  
in Fig.~\ref{fig2}. We also show for comparison  
the susceptibility of the resonant model, 

\begin{equation}
\label{resonant}
\chi_{T}= \frac{1}{2\pi^2 T} 
\phi\left(\frac{1}{2}+\beta\frac{\Delta}{4\pi}\right),
\end{equation}
where $\phi(z)=d^2 \ln\Gamma(z)/dz^2$ and $\Delta$ is 
the width of the resonant 
level. The latter has been determined by fitting  the data 
for $\tau_{0}\ T\le 0.2$ to 
Eq.~(\ref{resonant}) with the result $\Delta=0.827\  \tau_0^{-1}$. 
There is very good agreement between the theoretical 
expression and the Monte Carlo results in this 
temperature range. Deviations from the theoretical 
result 
are expected 
(and observed) at higher temperatures as Eq.~(\ref{resonant}) is 
only valid for $T\tau_0 \ll 1$. Taking the $T\to 0$ limit of 
Eq.~(\ref{resonant}) we find the zero-temperature susceptibility  
$\chi_{0}=2/(\pi\Delta) \approx 0.77\  \tau_0$. Defining  the Kondo 
temperature by  
$\chi_{0}=1/(2\ T^0_K)$, we obtain $T^0_K \tau_0 \approx 0.65$ .  

We have similarly computed the $\tau$-dependent susceptibility of the 
system for  several values of $J\ne 0$  
and $T$. The overall shape of the curves thus obtained is similar
 to that of 
those of  Fig.~\ref{fig1} but the decay of 
the correlations becomes slower and slower as $J$ increases. This is 
shown in Fig.~\ref{fig3} where we show results obtained for several 
values of $J$  at our lowest 
temperature, $T \tau_{0} = 32^{-1}$. 
This slowing down of the spin dynamics, which is accompanied of an 
increase of the susceptibility,  is a precursor effect of the phase 
transition that, as we shall see\  next, takes place for sufficiently large $J$. 

A necessary condition for the stability of the paramagnetic phase is that 
the inequality is $Y(T)\equiv 1-J\chi_{T}\ge 0$ holds \cite{bm}. $Y(T)$ 
 is plotted versus temperature in Fig.~\ref{fig4} for several values 
of $J$. The symbols are the Monte Carlo data. The dashed lines are 
fits of the results to a model that will be discussed 
below and that allows us to  extrapolate the results down to $T\to 0$.
We see that $Y(T=0)$ vanishes for 
$J=0.75\ \tau_{0}^{-1} \sim 1.15\  T_{K}^{0}$ which 
identifies it as the critical coupling. For $J > J_{c}$ $Y(T)$ vanishes at 
a finite temperature, $T_{g}$. 

Before discussing in detail the temperature 
dependence of the uniform susceptibility in the vicinity of the critical 
coupling, we 
shall make a disgression in order to  derive a 
simple model in terms of which the numerical  
data can be analyzed in a  transparent  way. 
We start by noticing that the frequency-dependent susceptibility 
may 
be related to the fluctuations of the auxiliary field $\xi(\tau)$. 
Using Eqs.~(\ref{singlespin}) and (\ref{sc2}) one may readily show that 

\begin{equation}
\label{xixi}
\left< \left|~\xi(i\omega_{n})~\right|^{2}\right >= Q(i\omega_{n}) 
\left[1 + \chi(i\omega_{n}) Q(i\omega_{n}) \right],
\end{equation}
where the expectation value on the lefthand side of the equation is 
taken 
with respect to the probability distribution 
${\cal P}[\xi(\tau)]\propto \exp(-\beta {\cal F})$ with 
 
\begin{equation}
\label{freexi}
{\cal F}[\xi(\tau)]
={\cal F}[0]+\frac{ kT }{2}\sum_{n}
\frac{|\xi(\omega_{n})|^{2}}{Q(\omega_{n})}
-kT \ln~\left<~{\cal T} 
\exp\left[~\int_{0}^{\beta}d\tau \xi(\tau) 
S_{z}(\tau)~\right]~\right>_{\Gamma}.
\end{equation}
Here, $\beta {\cal F}[0]=-\ln~\left[2\cosh\left(\beta 
\Gamma/2\right)\right]$ and 
$\left<~(\dots)~\right>_{\Gamma}=Tr\left[(\dots) \exp\left(\beta 
\Gamma S_{x}\right)\right]/Tr\left[\exp\left(\beta 
\Gamma S_{x}\right)\right]$. Assuming for the moment that the 
transverse part of the effective field dominates over its  
fluctuating longitudinal component, $\Gamma \gg 
~<\xi^{2}(\tau)>^{1/2}$, the free-energy (\ref{freexi}) may be 
expanded up to 
second order in $\xi$ :

\begin{equation}
\label{freexiap}
{\cal F}[\xi(\tau)]
={\cal F}[0]
+\frac{kT }{2}\sum_{n} 
\left[~Q^{-1}(\omega_{n}) - \chi^{(0)}(\omega_{n}) ~\right]
|\xi(\omega_{n})|^{2} + \dots,
\end{equation}
where the zeroth-order transverse susceptibility 
$\chi^{(0)}(\omega_{n})=\Gamma m_{x}/\left(\Gamma^{2}+
\omega^{2}_{n}\right)$ and $m_{x}=1/2 \tanh(\beta \Gamma/2)$.
Combining Eqs.~(\ref{freexiap}) and (\ref{xixi}) we derive  the 
following expression for the frequency-dependent susceptibility :
 
\begin{equation}
\label{chi}
\chi(\omega_{n})=\frac{1-\chi^{(0)}(\omega_{n}) 
~K(\omega_{n})-\sqrt{\left[~1 - \chi^{(0)}(\omega_{n}) 
~K(\omega_{n})~\right]^{2}-\left[~2 J \chi^{(0)}(\omega_{n}) 
~\right]^{2}}}{2 J^{2} \chi^{(0)}(\omega_{n})},
\end{equation}
where $K(\omega_{n})$ is the Fourier transform of the first term on 
the righthand side of Eq.~(\ref{v}). In the limit 
$T\tau_{0},~|\omega_{n}|\tau_{0} \ll 1$ this is 

\begin{equation}
\label{K}
K(\omega_{n})=\frac{2\pi\alpha}{\tau_{0}} (1 - |\omega_{n}| \tau_{0} 
+ 
\dots ).
\end{equation} 
Substituting this expansion in Eq.~(\ref{xixi}) we find :

\begin{equation}
\label{chifinal}
\chi(\omega_{n})=\frac{1}{J^{2}} \left\{ \frac{\omega_{n}^{2}}{\Gamma} + 
T_{K}+
\tilde{\alpha} |\omega_{n}| - \sqrt{ \left[ \frac{\omega_{n}^{2}}{\Gamma} + 
T_{K}+
\tilde{\alpha} |\omega_{n}| \right]^{2} - J^{2}} \right\}
\end{equation}
where $\tilde{\alpha}=\pi \alpha$, 
$T_{K}=\Gamma-\tilde{\alpha}/\tau_{0}$ 
and we have assumed $T \ll \Gamma$. From Eqs.~(\ref{chifinal}) and 
(\ref{xixi}) we can estimate $\left< \xi^{2}(\tau) \right> \approx \ 
(J^{2} + 2 \alpha/\tau^{2}_{0} )$ at large $\Gamma$. We therefore 
expect Eq.~(\ref{chifinal}) to be valid provided the condition 

\begin{equation}
\label{cond} 
\Gamma \tau_{0} \gg {\rm max}\{J\tau_{0} , \sqrt{2\alpha}\}, 
\end{equation}
is satisfied. It is clear that Eq.\ (\ref{cond}) will not be 
fulfilled by the bare 
parameters, in general. We can, however, imagine writing down a set of 
renormalization-group equations for the flow of the different couplings 
as the high energy cutoff is reduced. 
By analogy with the single-impurity case, we expect that  in the 
paramagnetic phase 
 the Kondo couplings will flow to the strong coupling fixed-point 
$\Gamma \equiv J^{K}_{\perp}\to \infty$. 
Therefore, we expect Eq.~(\ref{chifinal}) to become appropriate 
below some energy scale with renormalized values of the couplings. 

The derivation of the renormalization-group 
equations for the Kondo-alloy model is outside the scope of this work 
\cite{si}. We shall instead consider (\ref{chifinal}) as a 
phenomenological 
equation containing three renormalized parameters, $T_{K}$, 
$\tilde{\alpha}$, 
and $\Gamma$ to be determined by a fit of the numerical results.

The  ratio $T_{K}/J$ is determined by 
the static uniform susceptibility alone. Setting $\omega_{n}=0$ in 
Eq.~(\ref{chifinal}) 
we have :

\begin{equation}
\label{w-zero}
J\chi_{T}=\frac{T_{K}}{J}-\sqrt{\frac{T_{K}^{2}}{J^{2}}-1}.
\end{equation}
We thus see that the instability of the paramagnetic 
phase is 
signaled by the vanishing of the quantity under the square root in  
Eq.~(\ref{w-zero}). We may therefore take $\Delta=T_{K}^{2}/J^{2}-1$ 
as a measure of the distance to the quantum critical point and rewrite 
the susceptibility in the form :

\begin{equation}
\label{chidelta}
J\chi_{T}=\sqrt{1+\Delta} - \sqrt{\Delta}.
\end{equation}

Notice that, even if the assumptions made in the derivation 
of  Eq.~(\ref{chidelta}) are 
not valid, the latter can still be regarded as a 
parameterization of the susceptibility in terms of a new  
quantity,  $\Delta(T,J)$. The interest of this parameterization  
 stems from the fact that the $T$- and $J$-dependence of 
$\Delta$ is very simple. We show in Fig.~(\ref{fig5})  
the numerical values of $\Delta$ obtained inserting the  
Monte Carlo results for the static susceptibility
 in Eq.~(\ref{chidelta}). The dashed lines are fits to the 
 simple functional form :  

\begin{equation}
\label{fitdelta}
\Delta(T)=\Delta_{0} + (T/T_{0})^{3/2}, 
\end{equation}
where the parameters $\Delta_{0}$ and $T_{0}$ are functions of $J$ but 
not of $T$. The fits are very accurate over the entire 
temperature range of our simulations. The lowest curve, corresponding to 
our estimated value for the critical coupling,  
has been fitted with $\Delta_{0}=0$. Examination of the $J$ dependence of 
$\Delta_{0}$  shows that, near $J_{c}$, $\Delta_{0} \to a\  (1-J/J_{c})$,
 with $a \approx 1 $. The parameter $T_{0}$  has a finite limit,  
 $T_{0} \approx 0.27\  T_{K}^{0}$ as $J \to J_{c}$. 

Eqs.~(\ref{chidelta}) and (\ref{fitdelta}) imply that, in 
the neighborhood of the critical coupling, the uniform  
susceptibility has a non-Fermi-liquid $T$-dependence :

\begin{eqnarray}
\label{chidet}
J_{c} \chi_{T} \approx \left\{ 
\begin{array}{ll}
1 - \left(\frac{T}{T_{0}}\right)^{3/4} \ \ \ \ \  
& {\rm for}\ \ \ \ \  \Delta_{0}^{2/3} \ll T/T_{0} \ll 1, \\
\\
1- \sqrt{\Delta_{0}} - \frac{1}{2 \sqrt{\Delta{0}}} 
\left(\frac{T}{T_{0}}\right)^{3/2} \ \ \ \ \  
& {\rm for} \ \ \ \ \ T/T_{0} \ll \Delta_{0}^{2/3}.\\
\end{array}
\right.
\end{eqnarray}

The values of the exponent $\gamma$ found in the two regions 
 defined above 
correspond to those obtained by other authors 
\cite{sachdev1,sengupta,sachdev3} 
in the quantum critical (QC)  and quantum 
disordered I (QDI) regions  
in their analysis of metallic spin-glass models.
A third region (QDII) is expected at lower temperatures where the normal 
Fermi-liquid $T^{2}$ behavior is recovered. This second crossover is not 
visible in our data because, as we shall see, it occurs
  below the lowest temperature that we can reach in our
   quantum Monte Carlo simulations. 
  
The remaining two parameters in 
 Eq.~(\ref{chifinal}) may be determined from an analysis of 
 the full $\omega_n$- and $T$-dependence of the susceptibility. This is 
shown  in Fig.~\ref{fig6} for two couplings, $J=0.65\  J_{c}$ 
 and $J=J_{c}$. The symbols are
  the quantum Monte Carlo 
 data. The dashed lines are plots of Eq.~(\ref{chifinal}) with 
$\Gamma$,  $\tilde{\alpha}$ and $T_{K}$ adjusted  to 
fit the data. The quality of the fits
 is excellent  for 
all the values $J$ and $T$ considered. It is  remarkable 
that {\it all} our numerical results could be fitted 
with  the same values of 
$\Gamma \tau_{0}\approx 2.4$ and $\tilde{\alpha} \approx 1.48$. The 
full $J$- and $T$-dependence is therefore in the effective 
Kondo temperature, $T_{K}$. The values of  $T_{K}$ obtained 
from the fits of the frequency dependence of the 
susceptibility are  consistent with those determined above from 
the static susceptibility. 

At this point, we can  discuss the connection between our 
results and those of Sengupta and Georges \cite{sengupta}. 
Phrased in our language, their 
approximation for $\chi(\omega_{n})$ is obtained by ignoring the 
transverse field term in Eq.~(\ref{singlespin}), and solving 
the remaining  Ising  problem in the spherical approximation. The 
 expression for the dynamic susceptibility that results from this 
 procedure is :  

\begin{equation}
\label{seng-geor}
 \chi(\omega_{n}) \approx \frac{1}{J^{2}} \left\{ \lambda +
\tilde{\alpha} | \omega_{n} | - \sqrt{ \left[ \lambda+
\tilde{\alpha} | \omega_{n} | \right]^{2} - J^{2}} \right\}, 
\end{equation}
where $\lambda$ is a Lagrange multiplier introduced to impose the 
spherical constraint, 

\begin{equation}
\label{spherical}
\left<S_z(\tau) S_z(\tau) \right> = \frac{1}{\beta} 
\sum_{n= - \infty}^{\infty} 
\chi(\omega_{n})=1/4. 
\end{equation}
 As it stands,  expression (\ref{spherical}) diverges because  
Eq.~(\ref{seng-geor}) 
does not have the correct $\omega^{-2}_{n}$ high-frequency  behavior. 
A high energy cutoff $\Lambda$ must therefore be introduced in the 
simplified model. 

Comparison of Eqs.~(\ref{chifinal}) and (\ref{seng-geor}) shows that 
the two expressions become equivalent at low-frequencies  provided  
we identify $T_{K}$ with  $\lambda$ and $\tilde{\alpha} \Gamma$ 
with $\Lambda$. The fact that the Monte Carlo data could be fitted using 
Eq.~(\ref{chifinal}) with $T$- and $J$-independent 
values of  $\tilde{\alpha}$ and $\Gamma$ 
justifies {\it a posteriori} the use of a constant cutoff in the 
simplified model. Once this is fixed, the parameter 
$\lambda$  can be determined from condition (\ref{spherical}). Since  
the numerical data do satisfy this normalization, it is not surprising 
that the determination of $T_{K}$ from fits of the Monte Carlo data 
and that of $\lambda$ from enforcement of Eq.~(\ref{spherical}) 
result in the same  temperature dependence. 
This suggests to use Eq.~(\ref{chifinal}) in conjunction with
the normalization condition to estimate the $T$-dependence of $\Delta$ 
in the temperature region for which we do not posses numerical data. 
We show in Fig.~\ref{fig7} the result of applying this procedure 
to the case of $J=0.8\  J_{c}$.
We see that the expected crossover from a $T^{3/2}$ law to  normal $T^{2}$ 
behavior  occurs 
at a temperature $T^{\star} \approx 0.06\  T^{0}_{K}$. This is at the lower 
end of the temperature range that we can reach. The crossover temperature 
further diminishes as $J \to J_{c}$ where it vanishes. 
 This explains why normal behavior has not been seen in our simulations.

The imaginary part of the magnetic response can now be determined by 
analytic continuation of Eq.~(\ref{chifinal}). The general expression 
is complicated and not very illuminating. However, in 
the low frequency limit, $\omega \ll \tilde{\alpha}\  \Gamma$,
and for $J\to J_{c}$,  
Eq.~(\ref{chifinal}) can be cast in the scaling form :

\begin{equation}
\label{chi''}
J_{c} \chi''(\omega)=\sqrt{\Delta} \Phi\left(\frac{2 
\tilde{\alpha}\omega}{J_{c} \Delta}\right),
\end{equation}
where the universal scaling function $\Phi(x)$ 

\begin{equation}
\label{scalingfunc}
\Phi(x)=\frac{1}{\sqrt{2}} \ x \left[ \left(1+x^{2}\right)^{1/2} + 1 
\right]^{-1/2}. 
\end{equation}
This expression is equivalent to that found in references \onlinecite{sachdev1,sengupta,sachdev3}.

The low-frequency behavior of $\chi''(\omega)$ 
as $J \to J_{c}$ follows from the above equations :

\begin{eqnarray}
\label{loww}
J_{c} \chi''(\omega) \approx \left\{ 
\begin{array}{ll}
\frac{\tilde{\alpha} \omega}{J_{c}\Delta} \ \ \ \ \  
& {\rm for}\ \ \ \ \  \omega \ll J_{c}\Delta,\\
\\
\left(\frac{\tilde{\alpha} \omega}{J_{c}}\right)^{1/2}
& {\rm for} \ \ \ \ \ J_{c}\Delta \ll \omega \ll 
\tilde{\alpha} \Gamma.\\
\end{array}
\right.
\end{eqnarray}
The dissipative part of the susceptibility in the 
limit $\omega \to 0$ is Fermi-liquid-like 
everywhere except at the quantum critical point. However, the 
characteristic spin-fluctuation frequency 
$\omega_{\rm sf} \propto J_{c} 
\Delta$ vanishes as $J \to J_c$   
with $\omega_{\rm sf}\propto (1-J/J_{c})$ for $J\lesssim  J_{c}$
  and $\omega_{\rm sf} \propto 
T^{3/2}$ at the critical coupling.
The behavior of $\chi''(\omega)$ at $J_{c}$ is non-Fermi-liquid-like, 
$\chi''(\omega) \propto \sqrt{\omega}$ which reflects the slow 
 decay of the time-dependent spin-spin correlation function, 
$<S_{z}(t)S_{z}(0)> \sim t^{-3/2}$ that anticipates the 
appearence of long-range order in the system. The non-trivial 
$T$-dependence of the spin-fluctuation frequency in the vicinity of
 the quantum critical point that is 
responsible for the singular behavior of the susceptibility gives 
rise to anomalous powers in other thermodynamic and transport 
properties as well. In particular, Eqs.\ (\ref{chi''}) and 
(\ref{scalingfunc}) imply that the temperature corrections to
 the specific heat and the resistivity behave, respectively, as $\delta C/T \propto - \sqrt{T}$ and $\delta \rho \propto T^{3/2}$, in the 
quantum critical region \cite{sengupta}. 

The full $\omega$-dependence of the absorptive part of the dynamic susceptibility is shown in Fig.~( \ref{fig8}) for several temperatures 
at the critical coupling and several values of $J$ at $T=0.05\  T^0_K$.
 These curves have been computed by analytically continuing  
 the fits of the imaginary-frequency Monte Carlo data. 
These curves are very similar in shape to those obtained in reference \onlinecite{chakra} for the single-impurity Kondo model and may be
 characterized by an effective Kondo temperature that 
decreases with the distance to the quantum critical point where 
it vanishes. 
Indeed, Eqs.\ (\ref{chi''}) 
and (\ref{scalingfunc}) 
imply that the effective Kondo scale $\omega_{K}$, defined as the half-width of the 
relaxation function $\chi''(\omega)/\omega$, is $\omega_{K} \sim T_{K}^{(0)} \sqrt{\Delta}$.

\section{Conclusions}
\label{conclusions}

In this paper, we have studied numerically a Kondo lattice model with random 
exchange between localized spins. A mapping of this model to a 
self-consistent single-spin problem, exact in the limit of large 
lattice coordination, allowed us to  obtain a complete numerical
solution of the problem. 
The system has a quantum critical 
point  between a normal metal and a spin-glass state. There is a 
 region in the $T-J$ plane near the quantum critical point 
  where the characteristic 
 spin-fluctuation energy varies as a non-trivial power of temperature 
 ($\omega_{\rm sf} \propto T^{3/2}$). This gives rise to 
 non-Fermi-liquid behavior in thermodynamic and transport properties. 
At low enough temperature
normal Fermi liquid behavior is recovered, except at the critical 
coupling. Our numerical results can be very well described over a  large 
range of  frequency and temperature by  a simple model that we 
derive in the strong-coupling limit. This model is closely related to 
the $M$-component quantum-rotor and mean-field models that have been 
previously discussed 
in the literature \cite{sachdev1,sengupta,sachdev3}. 
Some interesting questions remain open, notably, to what extent
the assumption that the electronic bath remains unrenormalized
is a valid one. The enforcement of the self-consistency condition 
(\ref{g0loc}) poses some important technical difficulties which we hope 
to be able to overcome in future work.
One of us (M. J. R.) acknowledges support of Fundaci\'on Antorchas,
CONICET (PID $N^o4547/96$), and ANPCYT (PMT-PICT1855).

\begin{figure} 
\caption{Dynamic spin susceptibility in imaginary time for $J=0$ for 
 $\beta/\tau_{0}$= 18, 20, 24, 28 and 32, from top to bottom. 
The symbols are the Monte Carlo data. The solid lines are the fits 
referred to in the text.}
 \label{fig1} 
 \end{figure}

\begin{figure} 
\caption{Temperature dependence of the susceptibility for $J=0$. The 
circles are the Monte Carlo results. The dashed line represents the 
susceptibility of the resonant model.}
\label{fig2} 
\end{figure}

\begin{figure} 
\caption{The $J$-dependence of the dynamic spin susceptibility 
in imaginary time at  fixed temperature, $T\ \tau_{0} = 1/32 $. The 
values of $J\tau_{0}$ are $0, 0.6, 0.65, 0.7, 0.75$, from bottom to 
top. The symbols are the Monte Carlo data. The lines are guides for the eye.}
\label{fig3} 
\end{figure}

\begin{figure} 
\caption{The $J$- and $T$-dependence of $Y(J,T) = 1 - J \chi_{T}$. The 
paramagnetic phase is stable for $Y(J,T) > 0$.}
\label{fig4} 
\end{figure}

\begin{figure} 
\caption{The $J$- and $T$-dependence of $\Delta$. The symbols are the 
Monte Carlo data. The dashed lines are the fits to the expression 
in Eq.\ (\ref{fitdelta}).}
\label{fig5} 
\end{figure}

\begin{figure} 
\caption{The $\omega$- and $T$-dependence of the dynamic  
susceptibility for two values of the coupling $J$. The symbols are 
the Monte Carlo data. The lines are fits to the expression 
in Eq.\ (\ref{chifinal}).}
\label{fig6} 
\end{figure}

\begin{figure} 
\caption{The crossover between anomalous and  Fermi-liquid behavior 
 obtained from Eq.~(\ref{chifinal}) and condition (\ref{spherical}). } 
\label{fig7}
\end{figure}

\begin{figure} 
\caption{The relaxation function, $\chi''(\omega)/\omega$. 
The curves have been  obtained by 
analytic continuation of the  
 fits of the imaginary-time Monte Carlo data. (a) $T/T_{K} = 0.05$ and 
$J \tau_{0} =$ 0.55, 0.6, 0.65, 0.7 and 0.75, from top to bottom. (b) 
$J=J_{c}$ and $\beta/\tau_{0} = $ 32, 28, 24, 20, and 16, from top to 
bottom. }
\label{fig8} 
\end{figure}
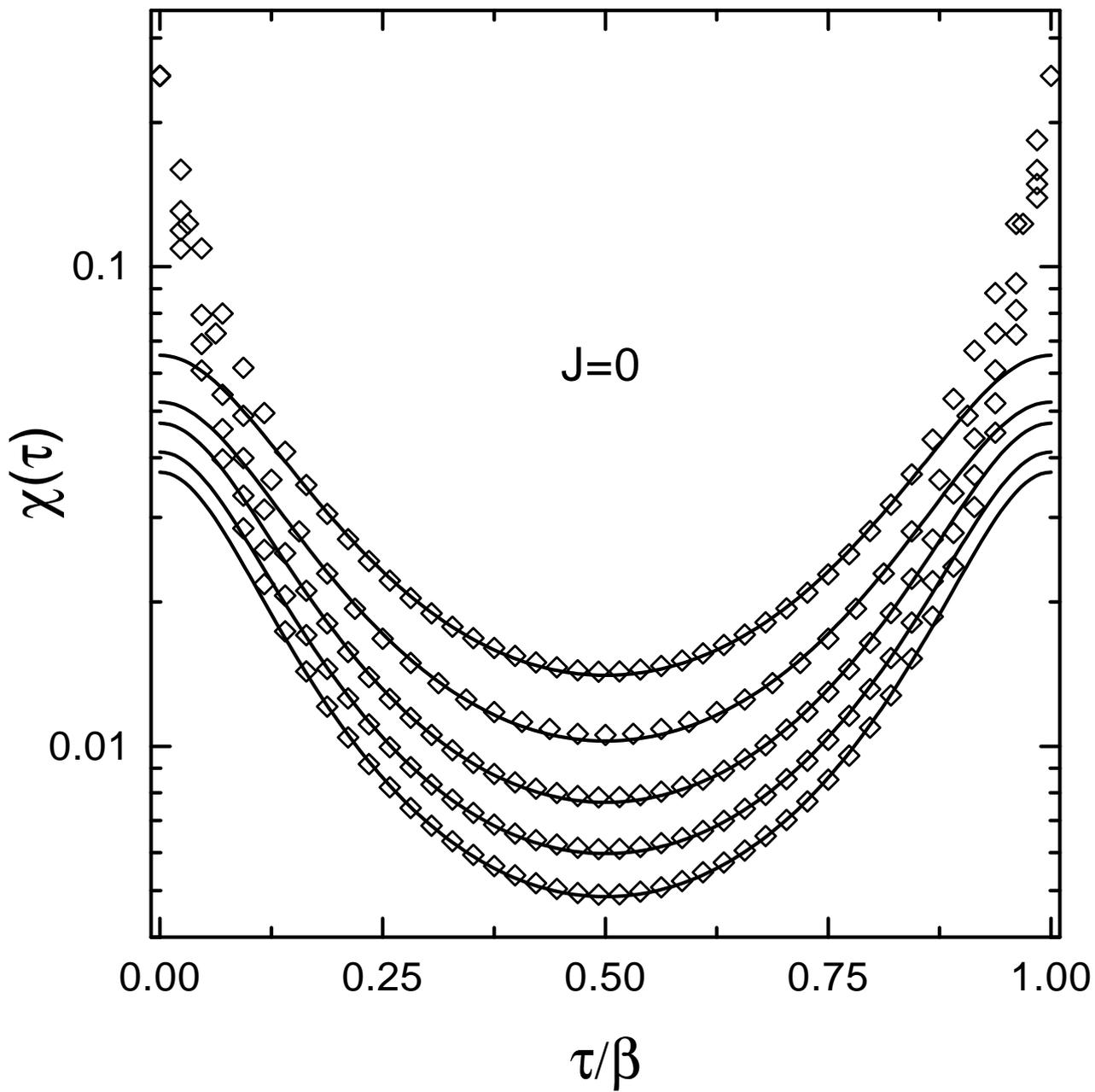
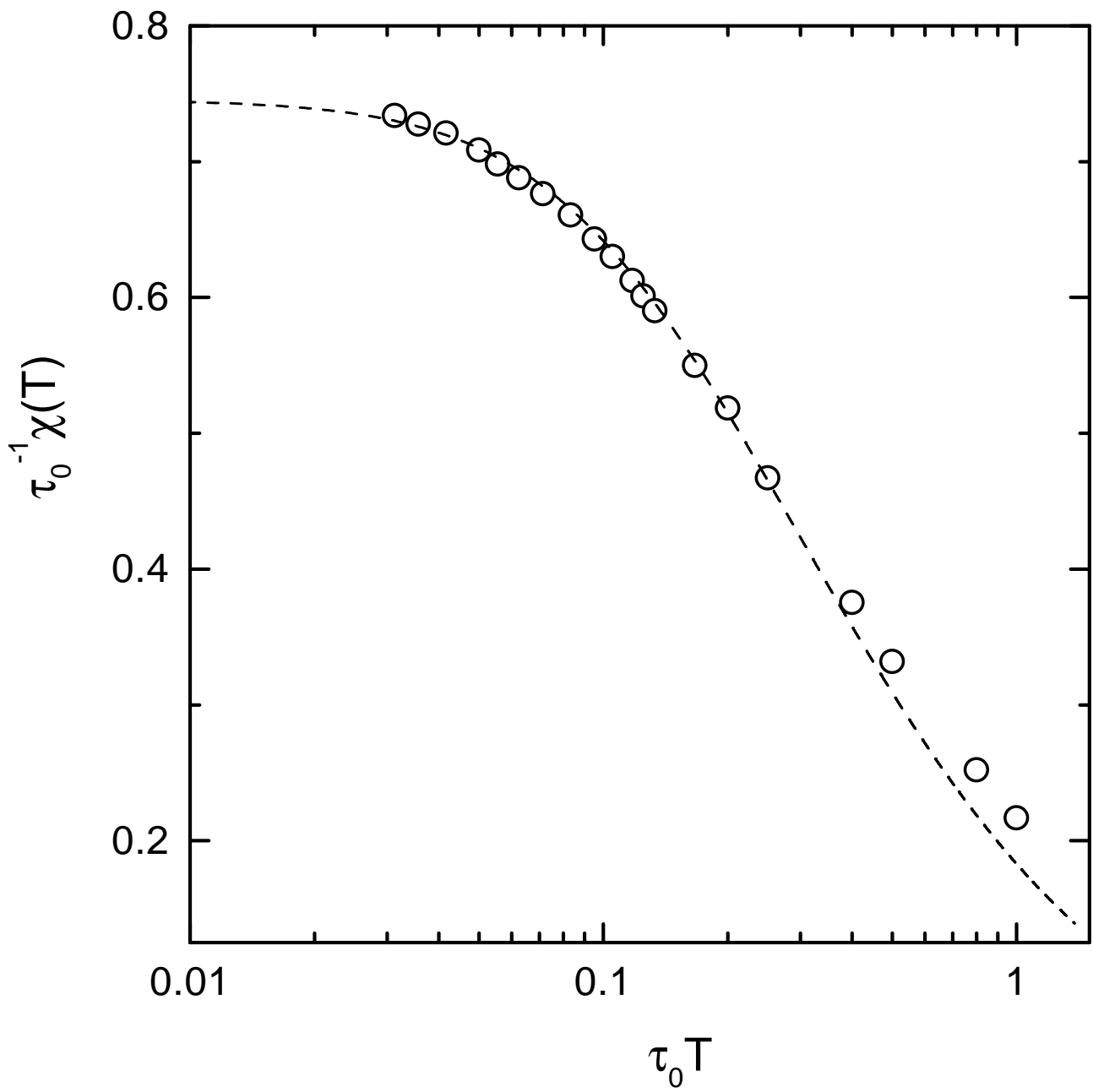
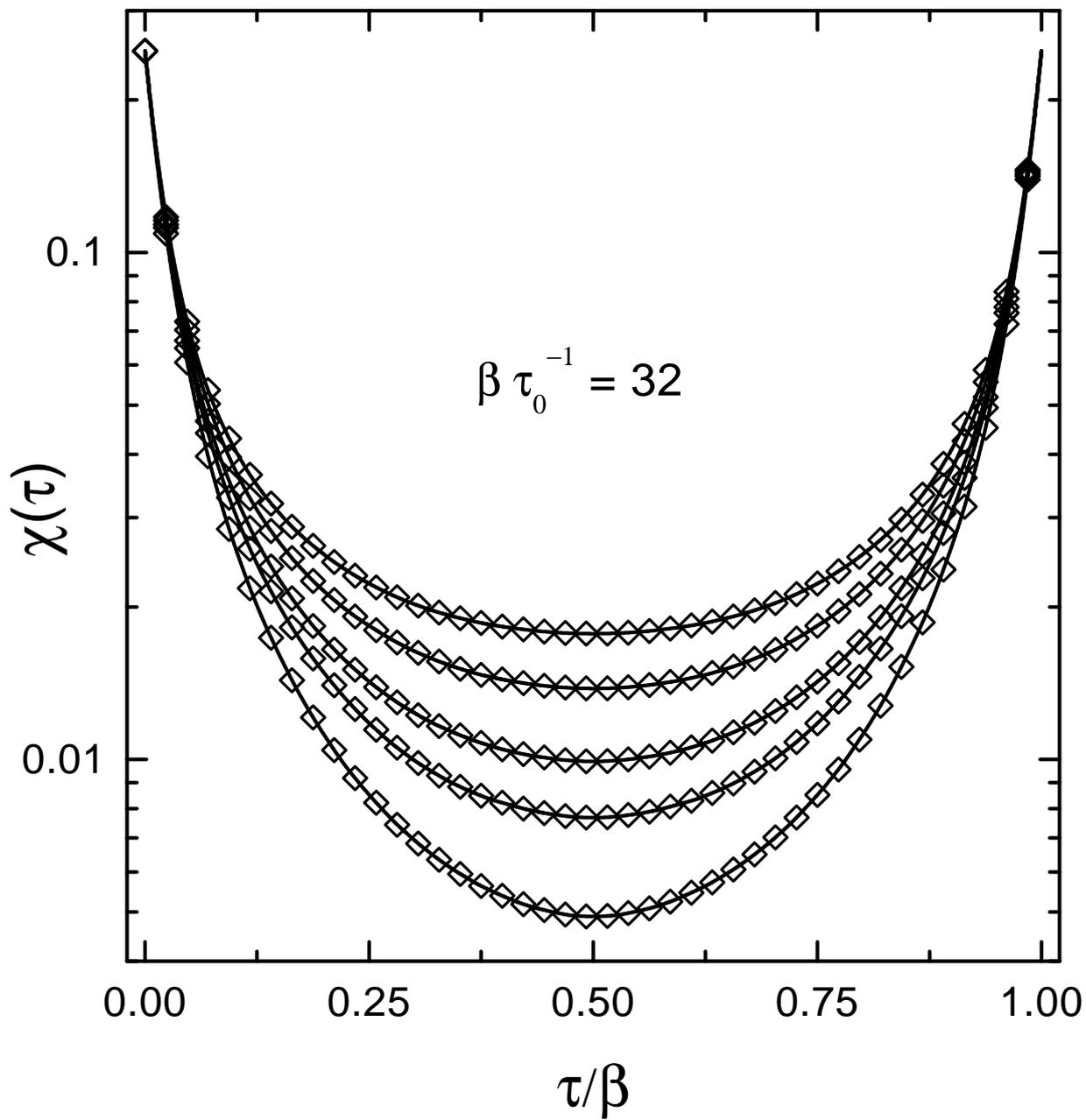
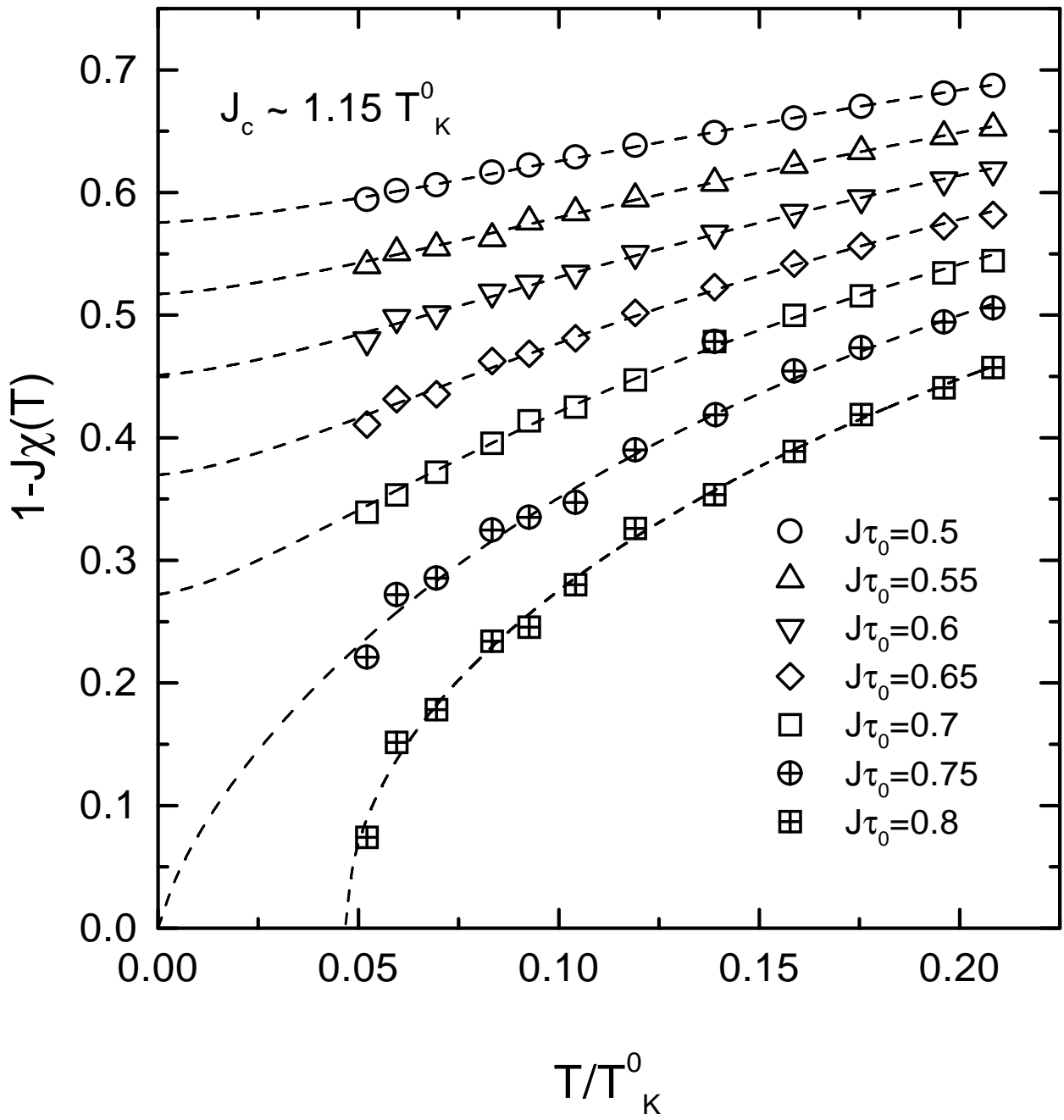
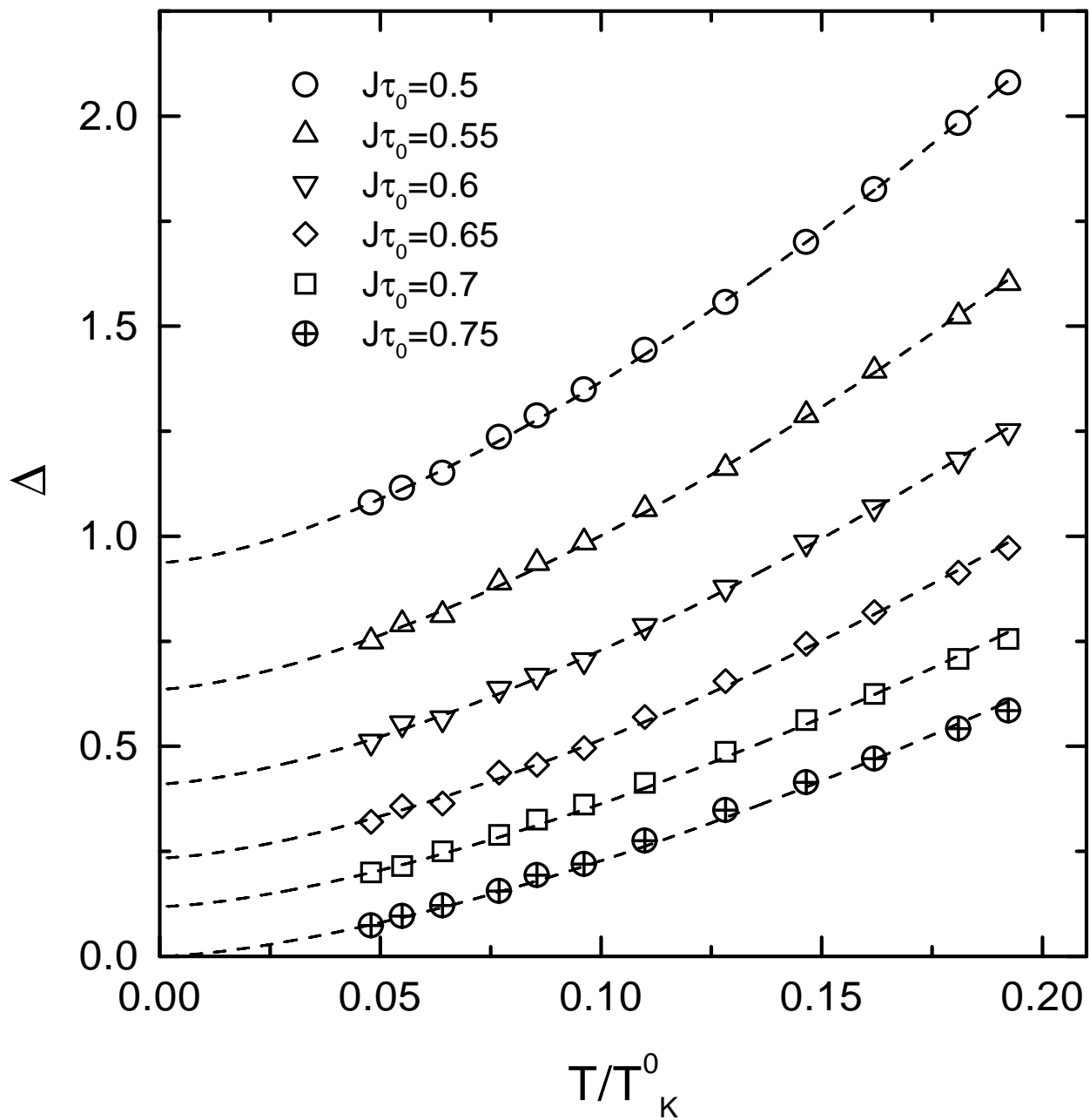
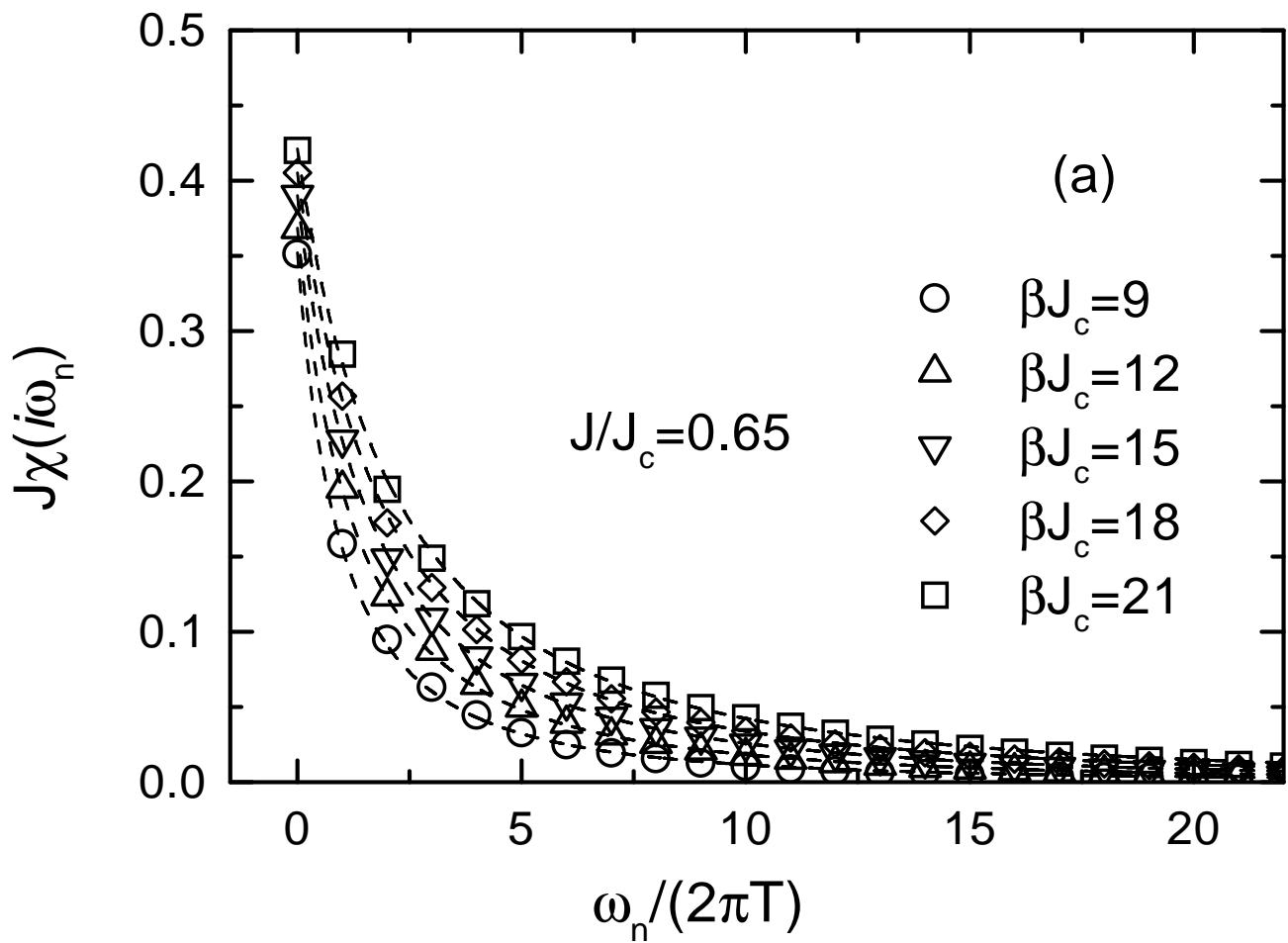
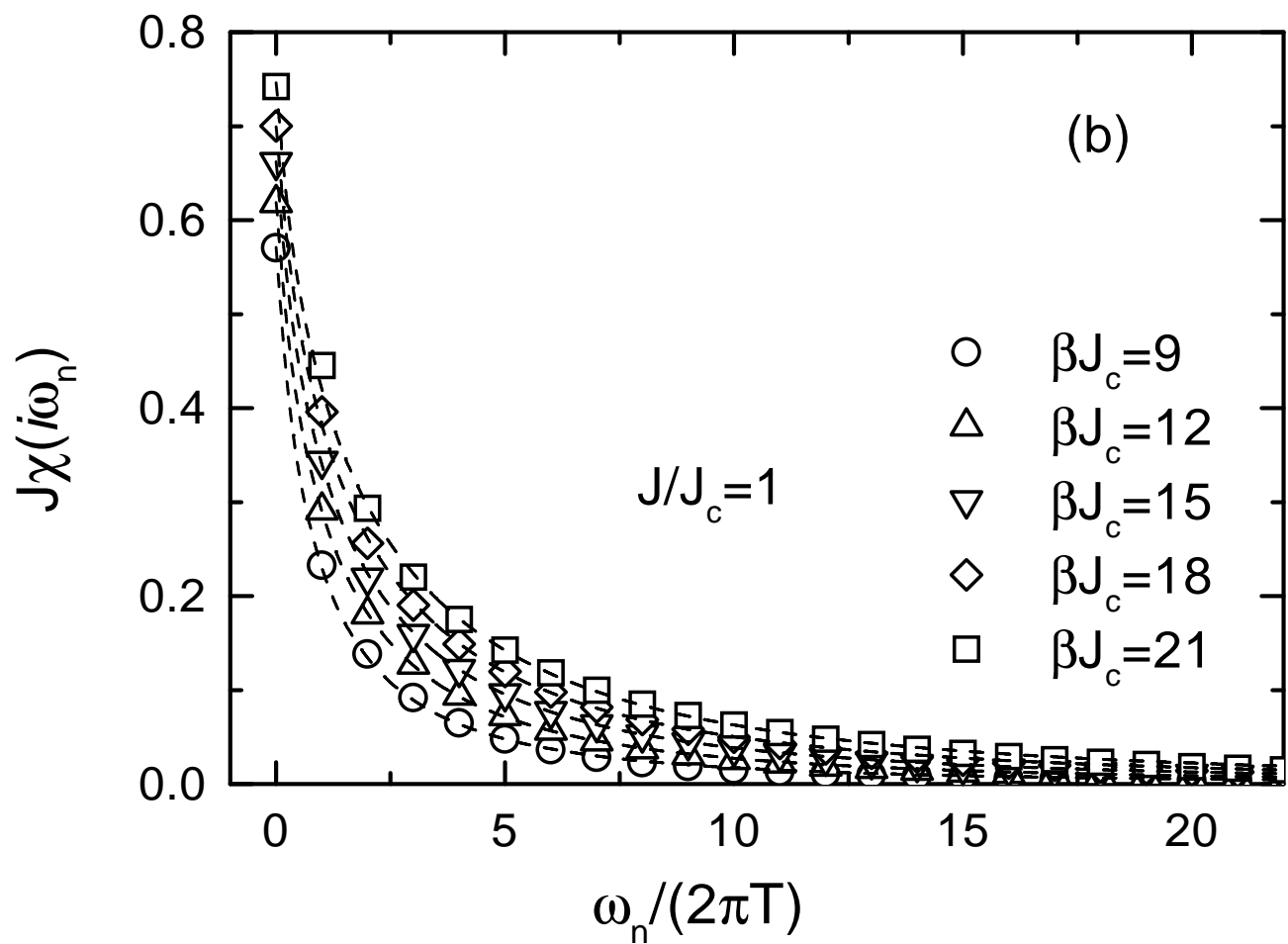
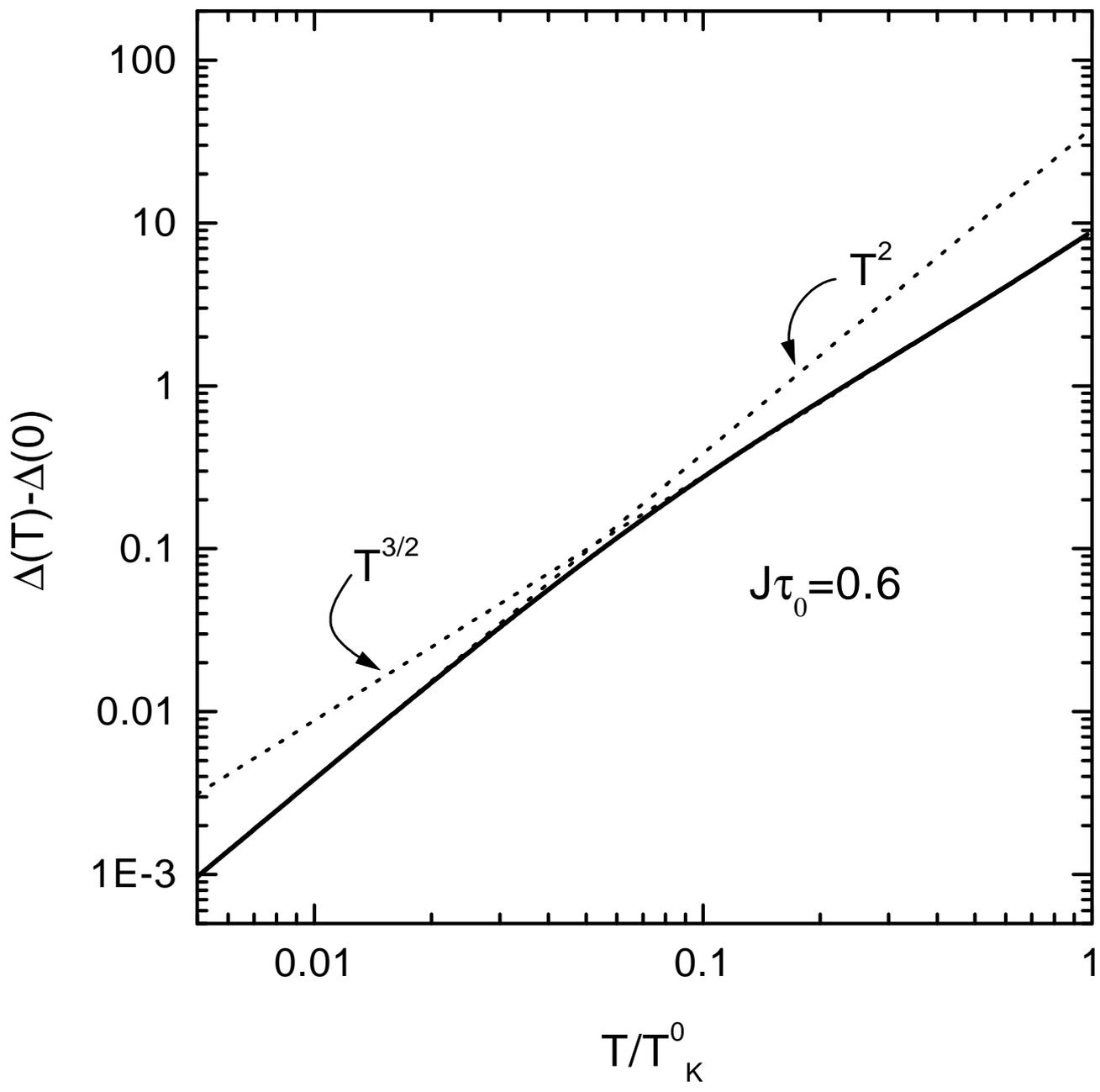
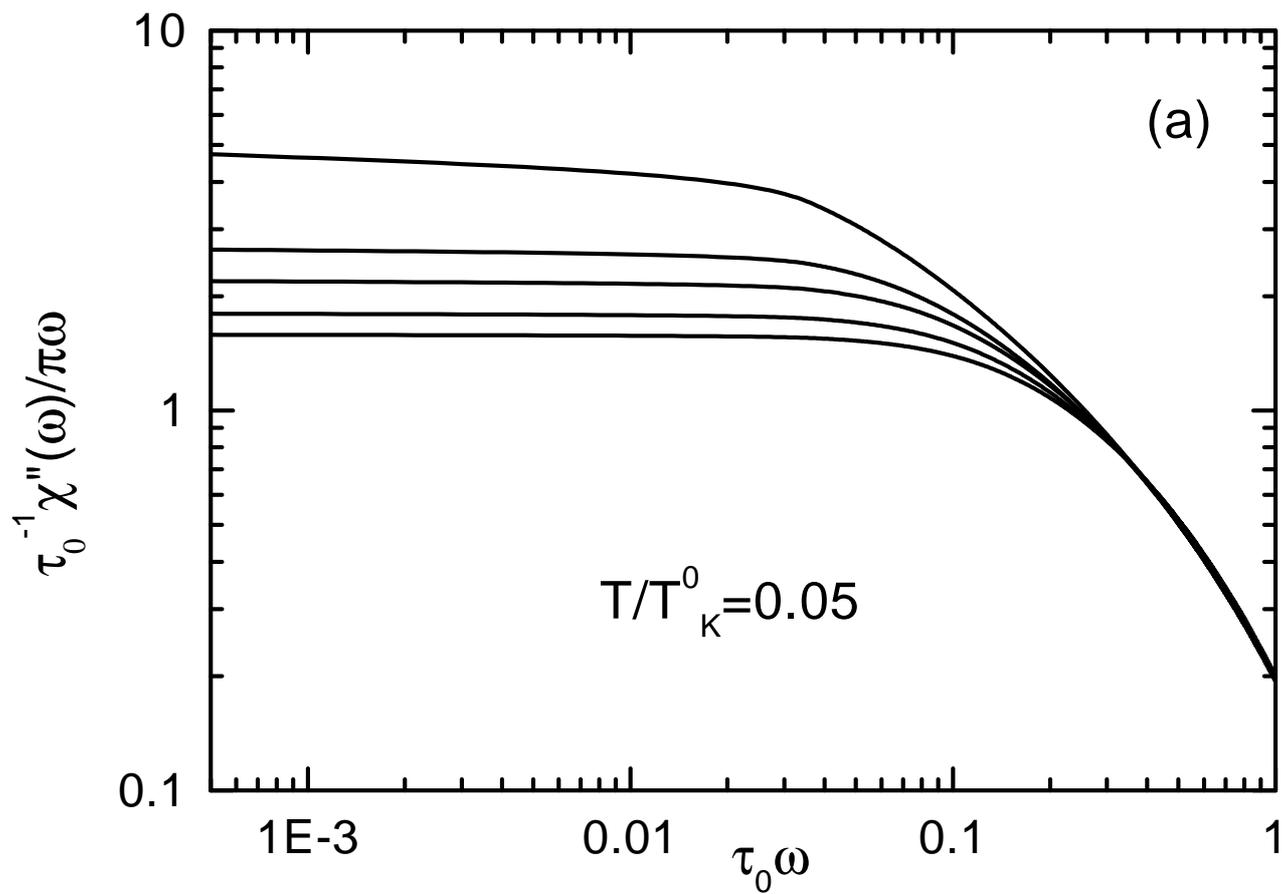
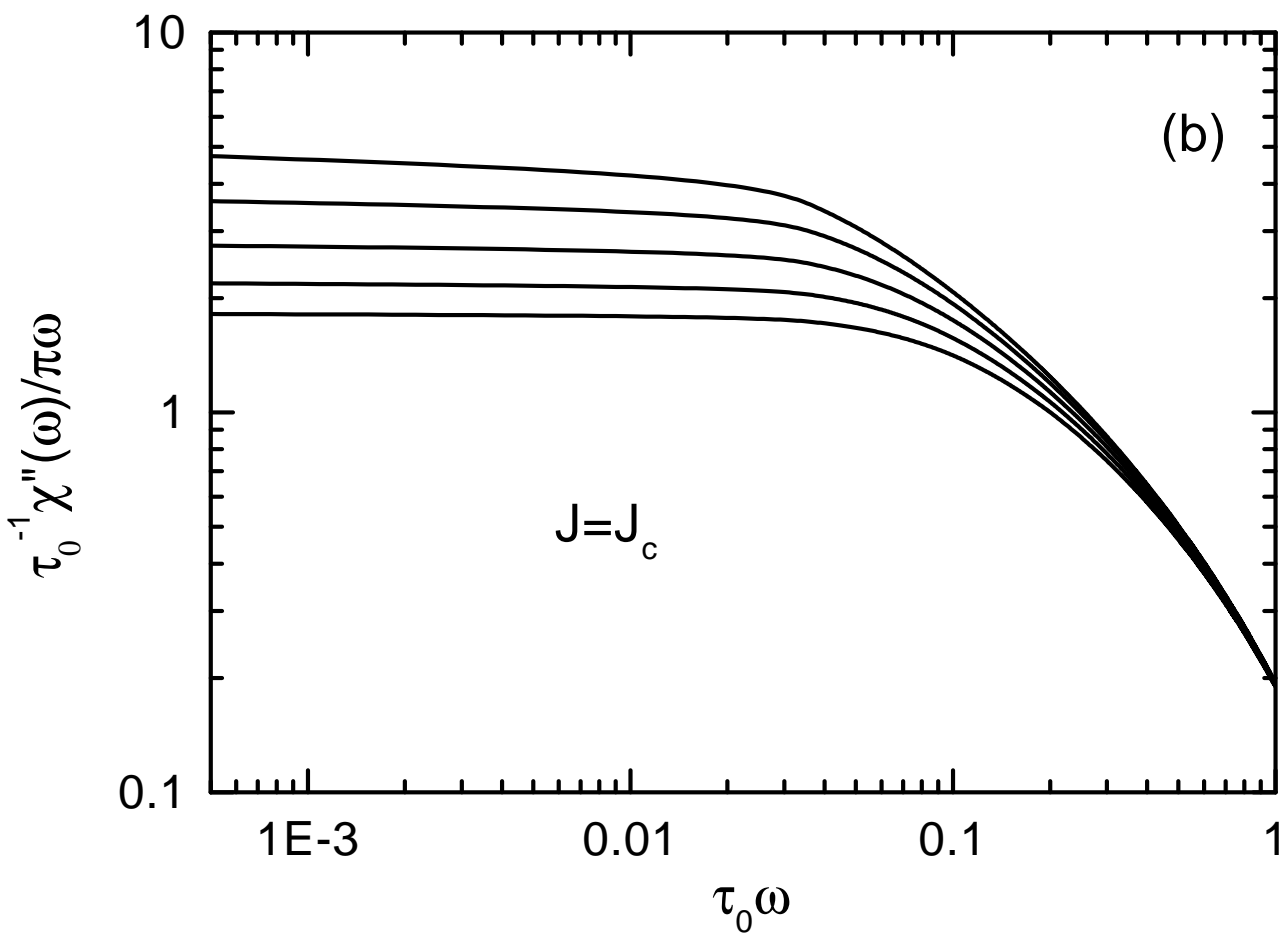
\end{document}